\documentclass[11pt]{iopart}
\usepackage[utf8]{inputenc}
\usepackage[T1]{fontenc}
\usepackage{graphicx}
\usepackage{amssymb}
\usepackage{nicefrac}
\usepackage{xcolor}
\usepackage{here}
\usepackage{tabularx}
\usepackage{bbm}
\usepackage[sort&compress,numbers]{natbib}

\usepackage{times}
\usepackage[colorinlistoftodos,prependcaption,textsize=small]{todonotes}

\usepackage{hyperref}

\hypersetup{
  colorlinks=true,
citecolor=black,
linkcolor=black,
urlcolor=black
  }

\newcommand{\bdt}[1]{\textcolor{black}{#1}}
\newcommand{\kct}[1]{\textcolor{black}{#1}}

\begin{document}

\title{Random acceleration process on finite intervals under stochastic restarting}

\author{Karol Capa{\l}a, Bart{\l}omiej Dybiec}
\address{Institute of Theoretical Physics,
and Mark Kac Center for Complex Systems Research, Jagiellonian University, ul. St. {\L}ojasiewicza 11, 30--348 Krak\'ow, Poland}
\ead{karol@th.if.uj.edu.pl, bartek@th.if.uj.edu.pl}

\date{\today}
\begin{abstract}
The escape of the randomly accelerated undamped particle from the finite interval under action of stochastic resetting is studied.
The motion of such a particle is described by the full Langevin equation and the particle is characterized by the position and velocity.
We compare three resetting protocols, which restarts \bdt{velocity or  position (partial resetting) and the whole motion (position and velocity -- full resetting).}
Using the mean first passage time we assess efficiency of restarting protocols in facilitating or suppressing the escape kinetics.
There are fundamental differences between \bdt{partial} resetting scenarios which restart velocity or position, as in the limit of very frequent resets only the position resetting (regardless of initial velocity) can trap the particle in the finite domain of motion.
The velocity resetting or the simultaneous position and velocity restarting provide a possibility of facilitating the undamped escape kinetics.

\end{abstract}

\pacs{
 05.40.Fb, 
 05.10.Gg, 
 02.50.-r, 
 02.50.Ey, 
 }

\maketitle

\section{Introduction}

Noise induced escape \cite{horsthemke1984,gardiner2009,redner2001} from a bounded domain \cite{benichou2005a} or a semi-bounded domain \cite{koren2007c} is the archetypal model underlying many  noise induced effects \cite{doering1992,mcnamara1989,gammaitoni1998,astumian1998,reimann2002,gammaitoni2009}.
Such an escape can be efficiently studied by stochastic methods \cite{horsthemke1984,gardiner2009}.
Evolution of single trajectories is described by the Langevin equation, which is the stochastic analogue of the Newton equation.
The Langevin equation \cite{coffey2012langevin} accounts for random perturbations, which are modelled by the noise term.
Typically, it is assumed that the noise is white and Gaussian, but multiple non-white or non-Gaussian extensions have been suggested \cite{klafter1987,klafter1996}.
The Langevin equation \cite{coffey2012langevin} is typically studied in two main regimes: the overdamped regime and in the regime of full (underdamped) dynamics.
In the underdamped regime, a random walker is characterized by position and velocity, while in the overdamped domain a random walker is fully characterized by its position only.
Therefore, the underdamped dynamics, although being more intuitive, allows for more options than the overdamped dynamics.

The Wiener process provides a mathematical description of the Brownian motion and it is one of the fundamental stochastic processes \cite{gardiner2009}.
It is the process described by the overdamped (first order) Langevin equation driven by the Gaussian white noise.
The generalization of the Wiener process to the regime of full (underdamped) dynamics is provided by the so-called
\bdt{random acceleration process \cite{bicout2000absorption} (because the acceleration is given by the noise term) which can be also called
the integrated Brownian motion (integrated Wiener process) \cite{barndorff2003integrated} or free inertial process \cite{masoliver1995exact}}.
In the regime of full dynamics, it is possible to study undamped \cite{goldman1971first,bicout2000absorption,masoliver1996exact,masoliver1995exact,lefebvre1989first,hesse2005first} and damped motions \cite{hintze2014small,hesse1991one,magdziarz2008fractional}.
Randomly accelerated motion of a free particle corresponds to the integrated Wiener process, while the damped motion to the integrated Ornstein--Uhlenbeck process \cite{lefebvre1989moment,duck1986first,lefebvre1989first,hesse1991one,hintze2014small,hesse2005first}.
Introduction of additional deterministic forces result in the forced motions \cite{srokowski2012anomalous,bai2017escape,capala2020underdamped}, but here we restrict ourselves to the undamped free randomly accelerated dynamics only.

The studied model \bdt{not only} extends examination of the exit time properties of the inertial equilibrium process driven by Gaussian white noise \cite{porra1994mean,masoliver1995exact,masoliver1996exact} by accounting stochastic resetting  \cite{evans2011diffusion,evans2011diffusion-jpa,evans2020stochastic}
\bdt{but also complements studies on properties of the unrestricted random acceleration process under stochastic resetting \cite{singh2020random}}.
Stochastic resetting is the protocol which can optimize the noise induced escape \cite{evans2011diffusion,evans2011diffusion-jpa,evans2020stochastic} by eliminating suboptimal trajectories.
The stochastic resetting is studied both theoretically \cite{reuveni2016optimal,debruyne2020optimization,ray2019peclet,ray2020space} and experimentally \cite{talfriedman2020experimental,besga2020optimal}.
The beneficial role of stochastic resetting is especially well visible in the case of escape from the half-line.
In contrast to escape from finite intervals, the escape from a half-line cannot be characterized by the mean first passage time, as it diverges.
The first passage time density is given by the L\'evy--Smirnoff distribution \cite{redner2001,samorodnitsky1994,chechkin2003b,koren2007,koren2007b} and it has power-law asymptotics with the exponent $-3/2$.
The asymptotics of the first passage time density is universal, as the same tail asymptotics is recorded for any symmetric Markovian drivings \cite{sparre1954,sparre1953,chechkin2003b,dybiec2016jpa}.
Under stochastic resetting the escape from the half-line is characterized by the finite mean first passage time \cite{evans2011diffusion}, because stochastic resetting prevents exploration of distant points.
Therefore, due to resetting, the particle cannot explore the whole space.
The same effect can be observed in escape from the potential well \cite{singh2020resetting,saed2019first} \bdt{due to narrowing of stationary states under resetting \cite{pal2015diffusion}. Therefore, the interpretation of the expedited escape kinetics from the potential well due to eliminating of suboptimal trajectories can be complemented by narrowing \cite{pal2015diffusion} of the quasi-stationary states \cite{vsiler2018diffusing,ryabov2019}.}
Optimization of the escape kinetics from the half-line \cite{evans2011diffusion} or from the potential well \cite{singh2020resetting} indicates that the stochastic resetting especially improves the kinetics when the domain of motion starts to resemble half-line \cite{dybiec2016jpa} or if it is unrestricted in the one of directions.
In such realms, resetting significantly decreases chances of visiting points which are far away from the absorbing boundary.

In the regime of the overdamped dynamics the particle motion is fully characterized by its position only.
Therefore, the restarting mechanism can affect the particle position only.
In the underdamped regime, there are more options of the stochastic resetting since randomly accelerated motion is characterized by the velocity and position, \bdt{but not all of them are of the renewal type \cite{bodrova2019nonrenewal,bodrova2019scaled,singh2020random}.}
Here, we study various resetting protocols which restart position, velocity or simultaneously both of them.
\bdt{The first two resetting protocols correspond to the so-called non-renewal (partial) resetting \cite{bodrova2019nonrenewal}, while the last scenario of full resetting is of the renewal type \cite{bodrova2019scaled}.}
\bdt{Within current studies,} we show fundamental differences and similarities between various restarting scenarios.

Here, we study \bdt{numerically} properties of \bdt{random acceleration process in bounded domains restricted by two absorbing boundaries accompanied by stochastic resetting}, i.e.,  the full (underdamped) stochastic dynamics of a free undamped inertial particle \bdt{in a confined geometry}.
The model under study is presented in the next section (Sec.~\ref{sec:model} -- Model).
Results of computer simulations are provided in Sec.~\ref{sec:results} (Results).
The paper is closed with Summary and Conclusions (Sec.~\ref{sec:summary}) \bdt{and accompanied by~\ref{sec:app}.}

\section{Model \label{sec:model}}

We study the escape of a free, inertial, undamped, randomly accelerated particle from the $(-l,l)$ interval restricted by two absorbing boundaries.
The motion is described by the underdamped (full) Langevin equation
\begin{equation}
m\ddot{x}(t)=  \sigma \xi (t),
\label{eq:langevind}
\end{equation}
where $\xi(t)$ is the Gaussian white noise satisfying $\langle \xi(t) \rangle=0$ and $\langle \xi(t) \xi(s) \rangle=  \delta(t-s)$.
\bdt{The parameter $\sigma$ defines the noise intensity while $m$ stands for the particle mass.}
\bdt{The process given by Eq.~(\ref{eq:langevind}) is the random acceleration process (RAP) \cite{theodore2014first}.}
Eq.~(\ref{eq:langevind}) can be transformed to the dimensionless \cite{capala2021inertial} variables $\tilde{x}$ and $\tilde{t}$
\begin{equation}
\left\{
\begin{array}{lcl}
\tilde{x} & = &  x/l, \\
\tilde{t} & = & t/T,
\end{array}
\right.
\label{eq:transformation}
\end{equation}
where
\begin{equation}
T=\left[  \frac{m l}{\sigma} \right]^{\frac{2}{3}}.
\label{eq:t}
\end{equation}
In such variables (after dropping tildes)
\begin{equation}
\ddot{x}(t)=  \xi (t),
\label{eq:langevin}
\end{equation}
and  $\langle \xi(t) \xi(s) \rangle=  \delta(t-s)$.
Consequently, there are no free parameters for the undamped free motion.
Moreover, absorbing boundaries are now located at $x=\pm 1$, i.e., the motion described by Eq.~(\ref{eq:langevin}) is studied as long as a particle stays within the $(-1,1)$ interval.

Randomly accelerated motion of a free particle, see Eq.~(\ref{eq:langevin}), corresponds to so-called the integrated Wiener process random acceleration process \cite{theodore2014first}.
Such a process constitutes a common research motif \cite{goldman1971first,lefebvre1989first,hesse2005first,tseng2004optimal,lindgren2008second,reymbaut2011convex,majumdar2010time,boutcheng2016occupation,kotsev2005randomly,burkhardt2007random}.
Among others, the escape of the integrated Wiener process from a bounded interval has been studied \cite{porra1994mean,masoliver1995exact,masoliver1996exact}.
Escape kinetics can be characterized by the mean first passage time (MFPT) which is the average time needed to leave the $(-1,1)$ interval for the first time
\begin{eqnarray}
\mathcal{T}(x_0,v_0) & = & \langle t_{\mathrm{fp}}(x_0,v_0) \rangle \\
&=&\langle\min\{t>0: x(0) = x_0 \wedge v(0)=v_0 \wedge |x(t)| \geqslant 1\}\rangle. \nonumber
\label{eq:mfpt-def}
\end{eqnarray}
The mean first passage time is the average of first passage times $t_{\mathrm{fp}}$.
From the ensemble of first passage times $t_{\mathrm{fp}}$ it is possible to calculate other escape characteristics, e.g., variance.
Since the randomly accelerated motion is characterized by the position and velocity, the mean first passage time depends both on the initial position $x_0$ ($x_0\in (-1,1)$) and the initial velocity $v_0$ ($v(0) \in \mathbb{R}$).
The general formula for the mean first passage, see \cite[Eq.~(17)]{masoliver1995exact} and \cite[Eq.~(4.3)]{masoliver1996exact}, significantly simplifies for the $v_0=0$ case
\begin{eqnarray}
\label{eq:integratedwiener-mfpt-dm}
\mathcal{T}(x,0) & = &
\frac{2}{3^{1/6} \Gamma(7/3)}
\left[ \frac{1+x}{2}  \right]^{1/6}
\left[ \frac{1-x}{2}  \right]^{1/6} \\
& \times &
\left\{
{}_2F\left(  1,-\frac{1}{3};\frac{7}{6}; \frac{1+x}{2}  \right)
+
{}_2F\left(  1,-\frac{1}{3};\frac{7}{6}; \frac{1-x}{2}  \right)
\right\} \nonumber,
\end{eqnarray}
where ${}_2F\left(  a,b ; c; x  \right)$ is the Gauss hypergeometric function \cite{abramowitz1964handbook}.

Within current studies, we extend the model of the escape of the free, inertial particle from the finite interval \cite{evans2011diffusion,evans2011diffusion-jpa,evans2020stochastic} by incorporating stochastic restarting, which is statistically independent of the underlying stochastic dynamics.
\bdt{The studied model extends RAP \cite{theodore2014first} considered in \cite{singh2020random} to finite domains and additional resetting protocols.}
In overdamped systems, i.e., systems characterized by the position only, the stochastic resetting is a protocol which at random time instants brings a particle back to a starting location $x_0$ \cite{evans2011diffusion,evans2011diffusion-jpa,evans2020stochastic}.
We apply the restarting with a fixed rate $r$ which corresponds to the so-called Poissonian resetting.
The duration of time intervals $\tau$ between two consecutive resets follow the exponential distribution, $\phi(\tau)=r \exp{(-r \tau)}$, where $r$ is the restarting rate.
In the case of full underdamped dynamics, not only position, but also the velocity can be reset.
Velocity resetting, as demonstrated in \cite{mandrysz2019energetics-pre} can suppress the unlimited energy growth \cite{mandrysz2020bounding}.
The position resetting brings a particle back to the initial position $x_0$, while the velocity restarting adjusts velocity back to $v_0$.
We consider three scenarios: (i) only the velocity is modified (the position is unaffected), (ii) the position is reset, while the velocity is unchanged and (iii) both the velocity and position are returned to initial values, i.e., the full process is restarted.
The motion described by Eq.~(\ref{eq:langevin}) is undamped, therefore velocity changes only due to action of the random force (noise) or stochastic resetting.


\section{Results\label{sec:results}}

We study three fixed rate restarting protocols:
(i) velocity, (Sec.~\ref{sec:resVelocity}),
(ii) position (Sec.~\ref{sec:resPosition})
and
(iii) position and velocity (Sec.~\ref{sec:resBoth}).
\bdt{Scenarios (i)  and (ii) correspond to partial (non-renewal) resetting \cite{bodrova2019nonrenewal}, while in (iii) the restarting is of renewal type \cite{bodrova2019scaled,singh2020random} as all variables are restarted.}
The duration of time intervals $\tau$ between two consecutive resets follow the exponential distribution with the rate parameter $r$.
The average time between two successive restarts is given by $\langle \tau \rangle = \frac{1}{r}$.

The role of resetting is studied \bdt{numerically} by the means of the mean first passage time.
MFPTs have been estimated from an ensemble of first passage times obtained by stochastic simulation of the Langevin equation (\ref{eq:langevin}).
The Langevin equation has been approximated by the Euler-Maruyama scheme \cite{higham2001algorithmic,mannella2002} with $\Delta t=10^{-4}$ and averaged over $N=10^3$ realizations.
Such a set of parameters provides a reasonable compromise between simulation time and accuracy of results.
Due to the system and dynamics symmetry, the mean first passage time satisfies the condition $\mathcal{T}(x_0,v_0)=\mathcal{T}(-x_0,-v_0)$.
\bdt{Consequently, in the special case of $v_0=0$ results are symmetric along $x_0=0$ line, while for $x_0=0$ the symmetry is along $v_0=0$ line.}
The mean first passage time, to a high extent, is determined by the initial velocity.
This is especially well visible for large $|v_0|$, when initial velocity favors motion into the direction of one of the absorbing boundaries.

For low resetting rate $r$, the mean first passage time is indistinguishable from the non-resetting case for all restarting protocols.
 \bdt{Consequently, for all resetting protocols, with $r\to0$, MFPT follows the general formula \cite[Eq.~(17)]{masoliver1995exact} and \cite[Eq.~(4.3)]{masoliver1996exact}. }
The difference arises for small but sufficiently large $r$.
In the opposite limit, of large $r$, the position resetting or the velocity resetting to $v_0=0$ can efficiently trap the particle in the system and induce the growth of MFPT as $r$ tends to infinity.
Therefore, there is a finite range of restarting rates which could optimize the escape process.
Furthermore, in order to keep clarity of presented figures, we do not show results for large $r$, because they can significantly extend the range of recorder MFPTs making figures illegible.
For completeness and verification of numerical simulations, results for \bdt{few selected $v_0$} with the low resetting rate have been compared with the exact value of the MFPT \cite{masoliver1995exact,masoliver1996exact}.
For $r\to 0 $, as expected, MFPT follows the theoretical curve, \bdt{e.g., for $v_0=0$} the one given by Eq.~(\ref{eq:integratedwiener-mfpt-dm}).
Importantly, as it will be demonstrated below, for appropriately selected initial conditions resetting can expedite the escape from the domain of motion.
The more detailed discussion, as being more case dependent, is provided in the following subsections.

%
\subsection{Restarting the velocity \label{sec:resVelocity}}

\begin{figure}[H]
     \centering
    \includegraphics[width=1\textwidth]{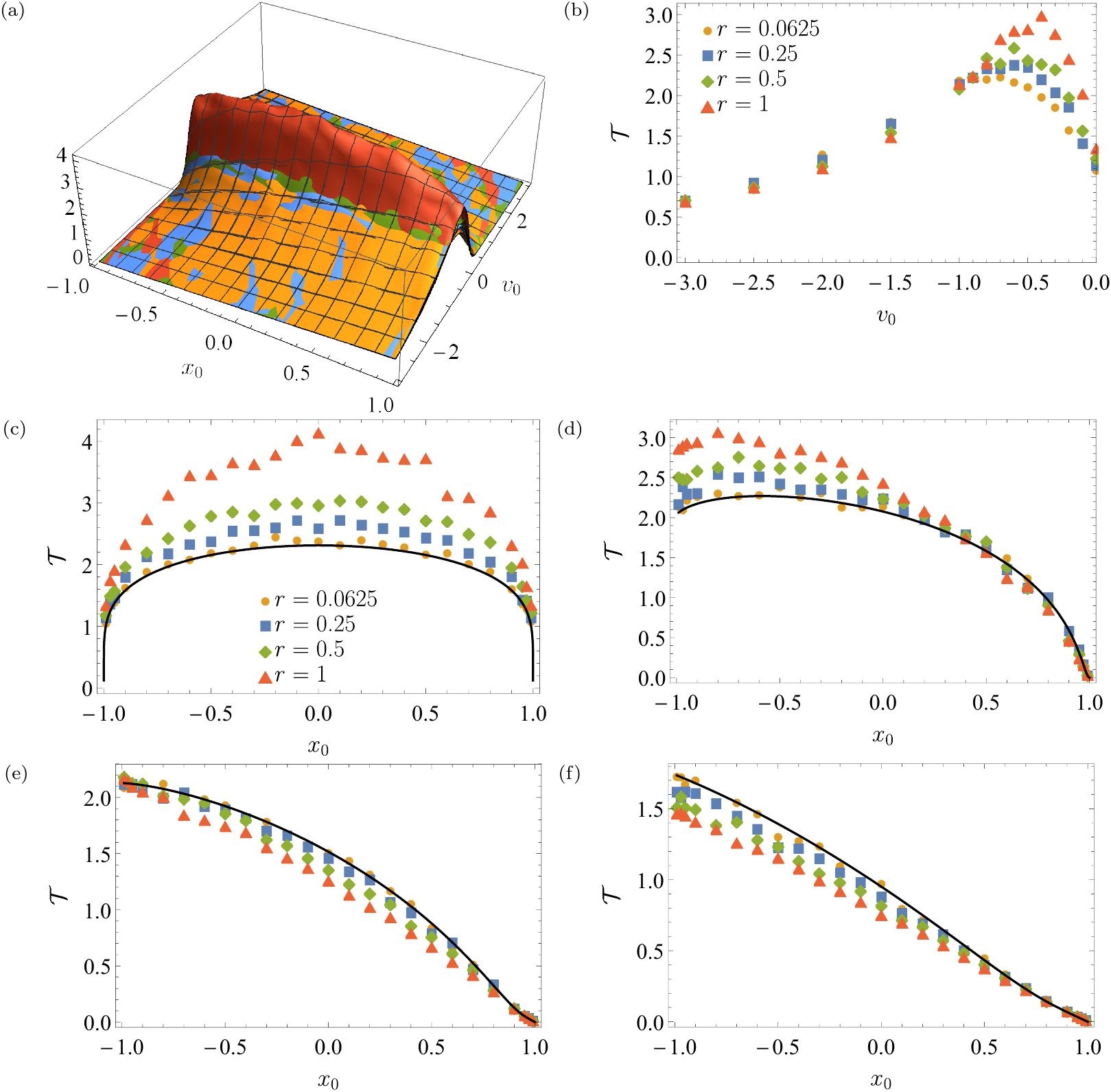}
    \caption{The MFPT $\mathcal{T}(x_0,v_0)$ as a function of the initial condition $(x_0,v_0)$ for various values of the velocity restarting rate $r$ (top left panel -- (a)).
    Cross-sections in the top right panel (b) correspond to the fixed $x_0$, i.e., $x_0=0.99$, while remaining cross-sections to the fixed initial velocity $v_0$:  $v_0=0$ (middle left panel -- (c)), $v_0=0.5$ (middle right panel -- (d)), $v_0=1$ (bottom left panel -- (e)) and $v_0=1.5$ (bottom right panel -- (f)). \bdt{Black solid lines in panels (c) -- (d) correspond to the exact formula for the MFPT, see Eq.~(\ref{eq:integratedwiener-mfpt-dm}) ($v_0=0$) and \cite[Eq.~(4.3)]{masoliver1996exact} ($v_0 \neq 0$).}
    }
    \label{fig:resVelocity}
\end{figure}

Fig.~\ref{fig:resVelocity}(a) shows dependence of the mean first passage time $\mathcal{T}(x_0,v_0)$ on initial conditions along with sample cross-sections for resetting in velocity.
\bdt{The MFPT in Fig.~\ref{fig:resVelocity}(a) is interpolated as a surface plot using the discrete grid of $(x_0,v_0)$ with
$x_0 \in\{ -0.99, -0.97, -0.95, -0.9, -0.8, \dots,  0.8, 0.9, 0.95, 0.97, 0.99\}$
and  
$v_0 \in  \{ -3.0, -2.5, -2.0, -1.5, -1.0, -0.9, -0.8, \dots,  0.8, 0.9, 1.0, 1.5, 2.0, 2.5, 3.0\}.$
}
The velocity restarting protocol, after random, exponentially distributed time, sets the instantaneous particle velocity to the initial velocity $v_0$ without affecting its position.
\bdt{Fig.~\ref{fig:resVelocity}(b) (top right panel of Fig.~\ref{fig:resVelocity})} depicts results for fixed initial position $x_0=0.99$, while remaining panels correspond to fixed initial velocities $v_0$:  $v_0=0$ (middle left -- panel (c)), $v_0=0.5$ (middle right -- panel (d)), $v_0=1$ (bottom left -- panel (e)) and $v_0=1.5$ (bottom right -- panel (f)).
\bdt{The examination of Fig.~\ref{fig:resVelocity}(a) shows that the $\mathcal{T}(x_0,v_0)=\mathcal{T}(-x_0,-v_0)$ symmetry, up to numerical accuracy, is preserved.}
Results for negative $v_0$, due to $\mathcal{T}(x_0,v_0)=\mathcal{T}(-x_0,-v_0)$ symmetry, can be reproduced from results with $v_0>0$ by reflecting MFPT curves along $x_0=0$.

As it is visible in Fig.~\ref{fig:resVelocity}(c), for $v_0=0$ resetting in the velocity slows down the escape, because it retains particle velocity while it does not change its position.
\bdt{Moreover, for $v_0=0$, the general symmetry of MFPT assures that $\mathcal{T}(x_0,v_0=0)=\mathcal{T}(-x_0,v_0=0)$, see Fig.~\ref{fig:resVelocity}(c)}.
The analogous effect \bdt{of slowing down the escape kinetics} is observed for small absolute values of initial velocity when (on average) restarting slows down the particle, because the velocity restarting can prevent escapes of particles starting in the vicinity of the boundary via the closest boundary by reverting the velocity towards the more distant boundary.
The opposite situation is recorded for $|v_0| \geqslant 1$, because in such a case restarting compels a particle to reach the boundary indicated by the initial velocity more efficiently.
Such a resetting can expedite escape regardless of the initial position $x_0$.
This effect is strongly connected to the fact that the escape of an inertial particle from a bounded domain is very sensitive to the initial velocity, especially for not too small initial velocities.
Moreover, the frequent restarting (moderate and large $r$) can be considered as the mechanism of maintaining constant velocity or at least as a mechanism of keeping the velocity distribution $p(v)$ narrow and centered around $v_0$.

In the limit of $r\to\infty$, restarting in velocity with $|v_0|>0$, keeps the particle velocity constant and equal to $v_0$.
Consequently, for $v_0\neq 0$, particles cannot be trapped in the $(-1,1)$ interval by restarting the velocity.
Moreover, the particle leave the domain of motion through the boundary in the direction of the initial velocity and the MFPT reads $\mathcal{T}(x_0,v_0)=\left|\frac{1-\mathrm{sign}(v_0)x_0}{v_0}\right|$ (results not shown).
\bdt{
Therefore, frequent velocity resetting resembles the overdamped, noise driven, escape from a finite interval under constant drift $k$ described by the overdamped Langevin equation
\begin{equation}
    \frac{dx}{dt}=k+\xi(t),
\end{equation}
for which the MFPT can be calculated from \cite[Eq.~(5.5.21)]{gardiner2009} and it is given by
\begin{equation}
 \mathcal{T}(x_0)= \frac{1+x_0 -2 e^{2k(1-x_0)} + (1-x_0) e^{4k}}{ \left( e^{4k}-1\right) k }.  
\end{equation}
For large $|k|$ the MFPT behaves like $\left|\frac{1-\mathrm{sign}(k)x_0}{k}\right|$, i.e., it is analogous to the MFPT under frequent velocity resetting.
Nevertheless, both setups bear fundamental differences.
In particular, for the full motion with frequent velocity resetting, escape takes only via the boundary indicated by the initial velocity.
For the overdamped motion, for not too large $|k|$ a particle can escape via any of two boundaries.
}
In contrast to $|v_0|>0$, for $v_0=0$, in the limit of $r\to\infty$ the particle does not leave the domain of motion and $\mathcal{T}\to\infty$ for every $x_0$.
However, this is the only one case when velocity resetting can completely prevent escape from the domain of motion.

\begin{figure}[H]
    \centering
    \includegraphics[width=1\textwidth]{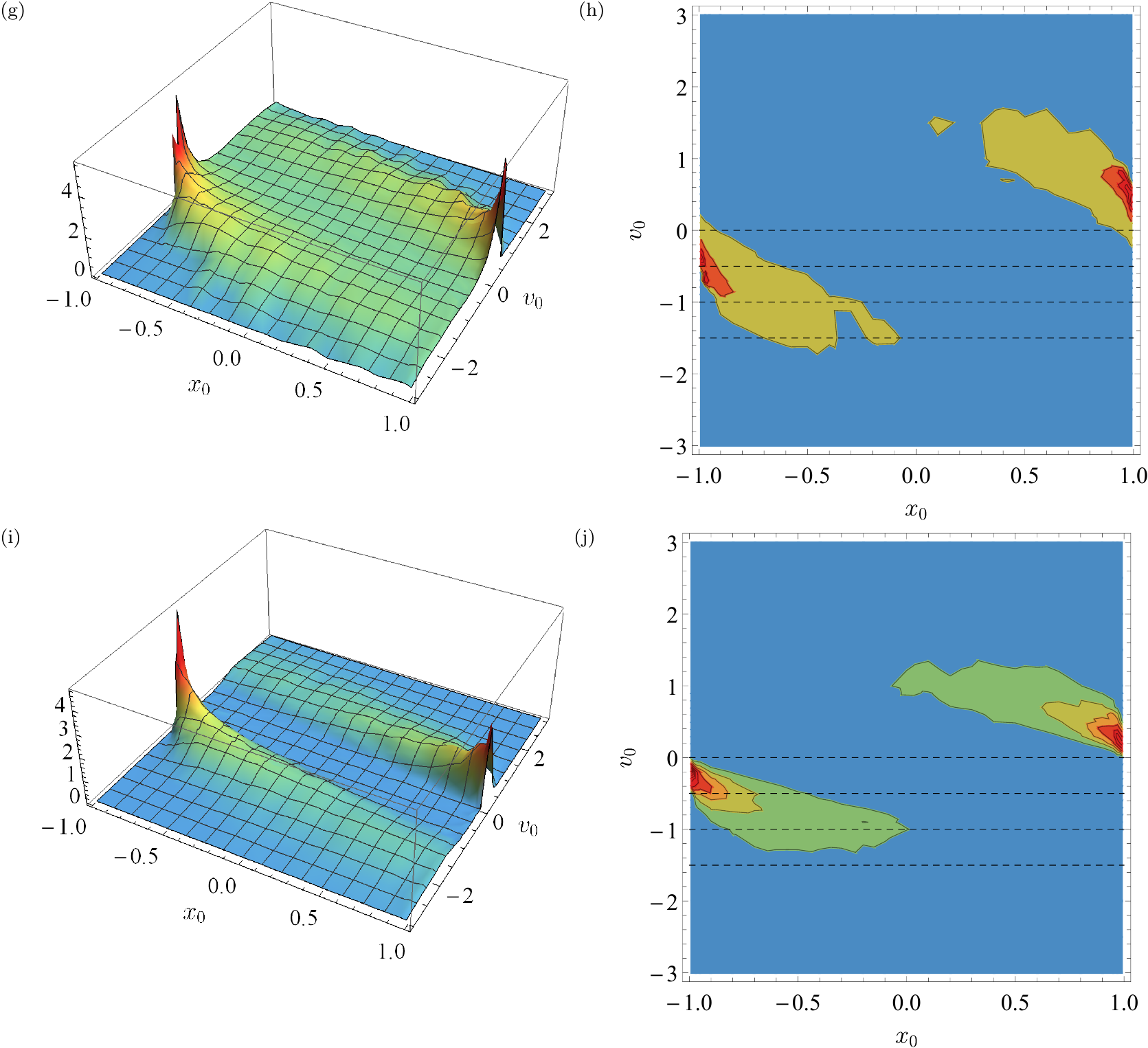}
    \caption{Surface and contour plots of the coefficient of variation $\mathrm{CV}$, see Eq.~(\ref{eq:cv}), (top row -- panels (a) and (b))  and resetting efficiency $\Lambda$, see Eq.~(\ref{eq:er}), (bottom row -- panels (c) and (d)) as a function of the initial condition $(x_0,v_0)$. In the top row, the dark blue region corresponds to $\mathrm{CV}<1$.
    Dashed horizontal black lines in right column (panel (b) and (d)) corresponds to MFPT cross-sections with $v_0\in\{0,0.5,1,1.5\}$, see Figs.~\ref{fig:CVxr0vr1}(c)--\ref{fig:CVxr0vr1}(d). }
    \label{fig:CVxr0vr1}
\end{figure}

The potential usefulness of stochastic resetting \bdt{for which renewal of all the degrees of freedom is retained}, i.e., the ability to accelerate the escape rate, can be recorded in domains where the coefficient of variation ($\mathrm{CV}$) is larger than one \cite{pal2017first}.
The coefficient of variation is the ratio between standard deviation of first passage times and the mean first passage time in the absence of stochastic resetting \cite{pal2019first}
\begin{equation}
    CV= \frac{\sigma(  t_{\mathrm{fp}})  }{ \langle t_{\mathrm{fp}} \rangle} = \frac{\sigma(  t_{\mathrm{fp}})}{\mathcal{T}}.
    \label{eq:cv}
\end{equation}
\bdt{It can be defined only in situations when MFPT and variance of first passage times are finite.}
Consequently, \bdt{despite the fact that velocity resetting is not of the renewal type}, along with numerical studies of MFPTs in undamped systems, see Fig.~\ref{fig:resVelocity}, we explore numerically calculated values of the coefficient of variation, see top row of Fig.~\ref{fig:CVxr0vr1}.
In the top row of Fig.~\ref{fig:CVxr0vr1} \bdt{(Figs.~\ref{fig:CVxr0vr1}(a) and~\ref{fig:CVxr0vr1}(b))} domains where $\mathrm{CV}>1$ are clearly visible.
Analogously to the symmetry of the mean first passage time, $\mathcal{T}(x_0,v_0)=\mathcal{T}(-x_0,-v_0)$, the very same symmetry holds for the coefficient of variation, i.e., $\mathrm{CV}(x_0,v_0)=\mathrm{CV}(-x_0,-v_0)$.
Maximal values of coefficient of variations are recorded for initial positions near absorbing boundaries with small initial velocities pointing to the closest absorbing boundary.
Maxima of the $\mathrm{CV}(x_0,v_0)$ surface are located within these domains because for a particle staring in the vicinity of the boundary with the small velocity pointing to this boundary the direction of motion can be easily reversed resulting in significant spreading of first passages times.
This in turn is responsible for the substantial increase in the coefficient of variation.

The stochastic resetting expedites the escape if it can eliminate suboptimal, i.e., very long trajectories.
Consequently, it has to restart trajectory to the point in the phase space from which it is harder to escape via a very long trajectory.
In the overdamped case, i.e., for the Wiener process, as indicated in \cite{pal2019first}, the stochastic resetting does not expedite the escape from the $(-1,1)$ interval  if a particle is reset to $|x_0| < \frac{1}{\sqrt{5}}$.
Restarting motion to $x_0$ such that $|x_0| < \frac{1}{\sqrt{5}}$ increases the MFPT by increasing the likelihood of emergence of suboptimal trajectories.
Stochastic resetting facilitates the escape kinetics when a particle restarts its motion from a point which is not too distant from the boundary.
\bdt{Restarting this way} decreases chances of escaping via a more distant boundary, therefore with moving a restarting point closer to the boundary an escape from the finite interval starts to resemble the escape from the half-line, which is significantly modified by resetting \cite{evans2011diffusion}.
In the underdamped case the situation is more complex, because not only the initial position but also the initial velocity determines the mean first passage time.
The larger the initial velocity is, the more dominating role it plays.

The ability of stochastic resetting to facilitate the escape kinetics can be discriminated by the resetting efficiency $\Lambda$
\begin{equation}
    \Lambda(x_0,v_0)=\frac{\mathcal{T}(x_0,v_0,r=0)}{\min_{r} \mathcal{T}(x_0,v_0,r) }-1,
    \label{eq:er}
\end{equation}
which is the ratio of the without resetting MFPT and the minimal MFPT under resetting recorded for the fixed initial condition.
Resetting can expedite the escape kinetics if $\Lambda(x_0,v_0)>0$.
Therefore, the examination of $\Lambda(x_0,v_0)$ allows us to assess if for a given resetting protocol and initial conditions restarting can increase the escape rate.
The dependence of the resetting efficiency $\Lambda$ on the initial condition $(x_0,v_0)$ is depicted in the bottom row (panels (c) and (d)) of Fig.~\ref{fig:CVxr0vr1}.
The resetting efficiency inherits symmetries of mean first passage time and coefficient of variation, i.e., $\Lambda(x_0,v_0)=\Lambda(-x_0,-v_0)$.
The examination of the efficiency of resetting reveals that velocity restarting can indeed expedite the escape from the interval restricted by two absorbing boundaries.
The domain in which $\Lambda>0$ partially overlaps the $\mathrm{CV}>1$ region, but it is not the same.
This indicates that, \bdt{due to non-renewal character of partial resetting}, the condition $\mathrm{CV}>1$ gives only some proxy for the domain where resetting can facilitate the escape kinetics.

\begin{figure}[H]
    \centering
    \includegraphics[width=0.7\textwidth]{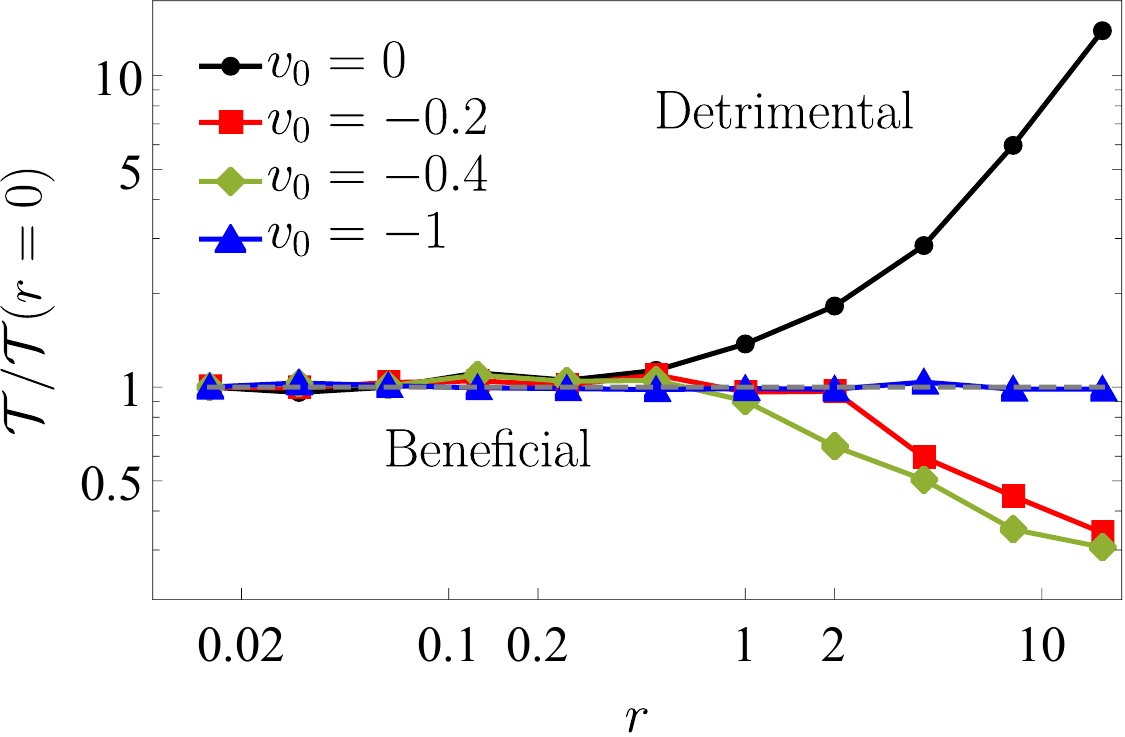}
    \caption{The ratio $\mathcal{R}=\mathcal {T}/\mathcal{T}(r=0)$ as the function of the resetting rate $r$ for the fixed initial position $x_0=-0.95$ and various initial velocities $v_0$.
    \bdt{In order to show the boundary between beneficial and detrimental domains the separating line $\mathcal{T}(r)/\mathcal{T}(r=0)=1$ along with labels have been added.}
    }
    \label{fig:r-xr0vr1}
\end{figure}

Fig.~\ref{fig:r-xr0vr1} presents individual $\mathcal{T}(r)$ curves divided by $\mathcal{T}(r=0)$ corresponding to the fixed initial position $x_0=-0.95$ and various initial velocities $v_0$.
\bdt{The ratio $\mathcal{R}=\mathcal{T}(r)/\mathcal{T}(r=0)$, indicates whether restarting is beneficial ($\mathcal{R}<1$) or detrimental ($\mathcal{R}>1$).}
In accordance with earlier observations, for $v_0=0$ resetting inevitably results in an increase of the MFPT while for $v_0=-1$ the MFPT is insensitive to the resetting.
In the intermediate regime, the restarting facilitates the escape kinetics because frequent resetting maintains the constant velocity.
Consequently, minimal MFPTs are observed for large values of resetting rate $r$.

%
\subsection{Resetting in position \label{sec:resPosition}}

\begin{figure}[H]
    \centering
    \includegraphics[width=1\textwidth]{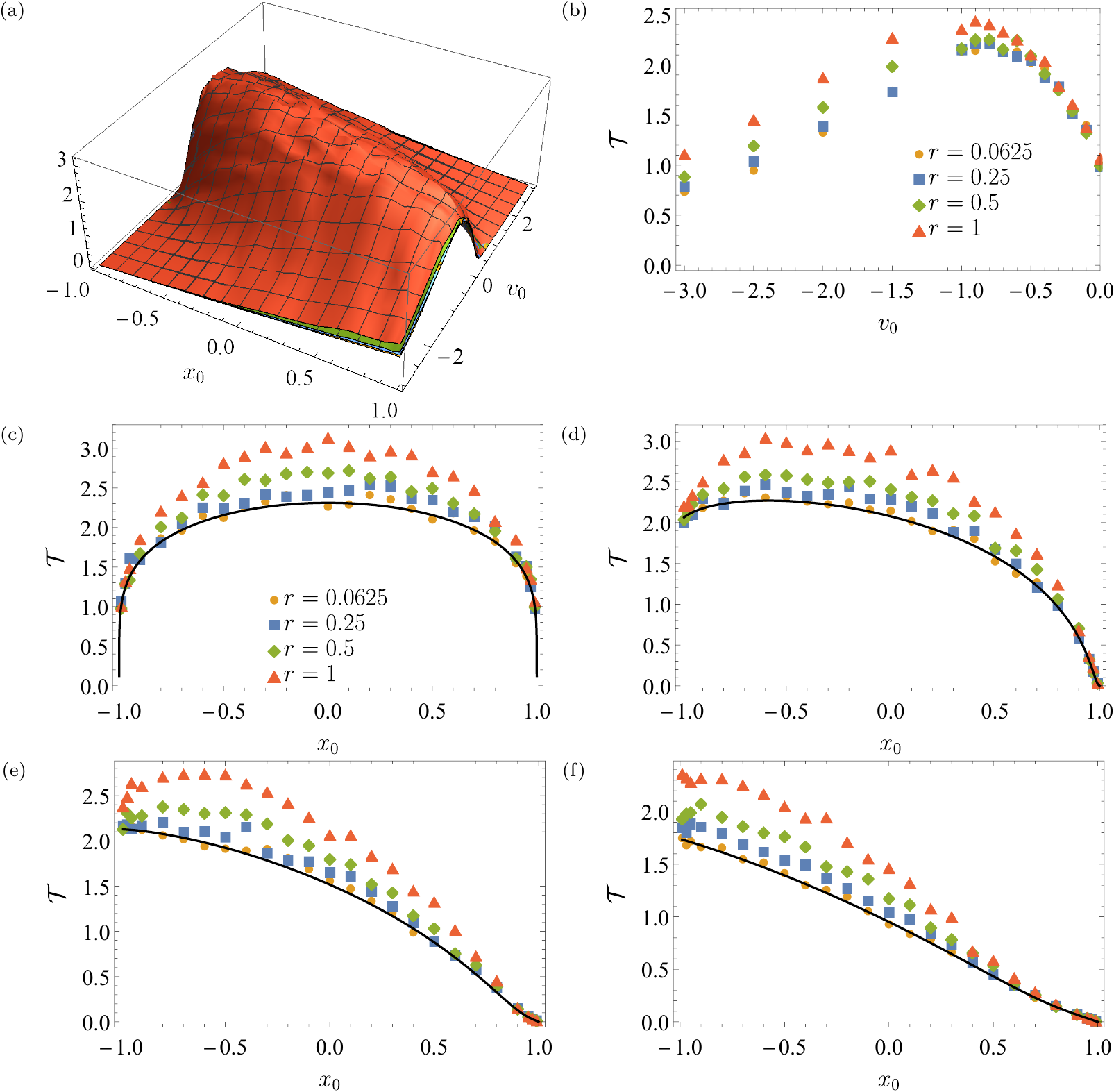}
    \caption{The same as in Fig.~\ref{fig:resVelocity} \bdt{(dependence of the MFPT on initial conditions and resetting rate)} for position resetting.
    }
    \label{fig:resPosition}
\end{figure}

Fig.~\ref{fig:resPosition}, in an analogous manner  like  Fig.~\ref{fig:resVelocity}, shows the dependence of the mean first passage time $\mathcal{T}(x_0,v_0)$ on initial conditions along with sample cross-sections for the \bdt{partial (non-renewal)} resetting in the position.

In comparison to the velocity restarting, see Sec.~\ref{sec:resVelocity}, for the same set of parameters, the position resetting slows down the escape less than velocity resetting.
This becomes immediately evident from comparison of Fig.~\ref{fig:resVelocity} and Fig.~\ref{fig:resPosition}, as in Fig.~\ref{fig:resPosition} MFPT attains smaller values than in Fig.~\ref{fig:resVelocity}.
Moreover, for $|v_0| > 0.5$, the order of individual curves in cross-sections is different than in Fig.~\ref{fig:resVelocity}.
Resetting in position does not modify the particle velocity but it moves a particle to the initial location, which can be less beneficial than a current position.
For instance, especially for the large initial velocity and remote initial position, a particle can fast approach the absorbing boundary, but resetting can move the particle back to $x_0$ which is distant from the approached boundary.

In contrast to the velocity restarting, the position resetting for any $x_0$ from the interval of motion is capable of trapping the particle in the domain of motion.
Such a trapping is recorded for $r \to \infty$
Therefore, for sufficiently large resetting rates MFPT becomes larger than for the velocity resetting.
Another interesting effect is associated with the velocity distribution.
The position resetting brings a particle back to its initial position without changing its velocity.
Therefore, the $p(v)$ distribution for all times $t$ is given by the Gaussian distribution with the mean value determined by the initial velocity and growing (linearly) in time variance.
Thanks to increasing width of the velocity distribution, the particle's velocity and position fluctuate more, increasing chances of escaping from the domain of motion.
These chances are decreased by the position resetting which statistically moves the particle away from the boundaries of the domain of motion resulting in the increase of the MFPT.

\begin{figure}[H]
\centering
    \includegraphics[width=1\textwidth]{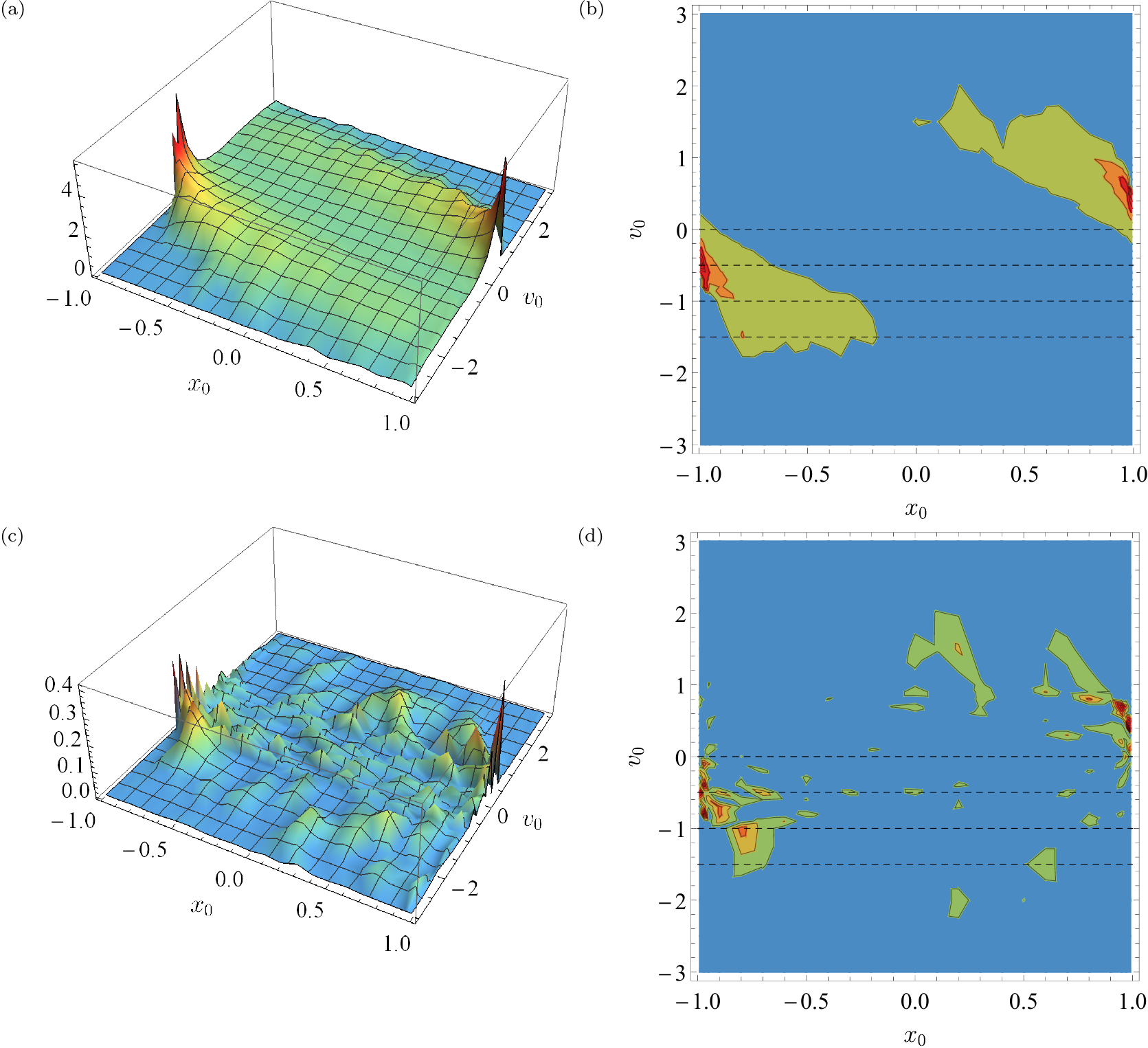}
    \caption{The same as in Fig.~\ref{fig:CVxr0vr1} \bdt{(dependence of $\mathrm{CV}$ (top row) and $\Lambda$ (bottom row) on initial conditions)} for position resetting.}
    \label{fig:CVxr1vr0}
\end{figure}

\begin{figure}[H]
    \centering
    \includegraphics[width=0.7\textwidth]{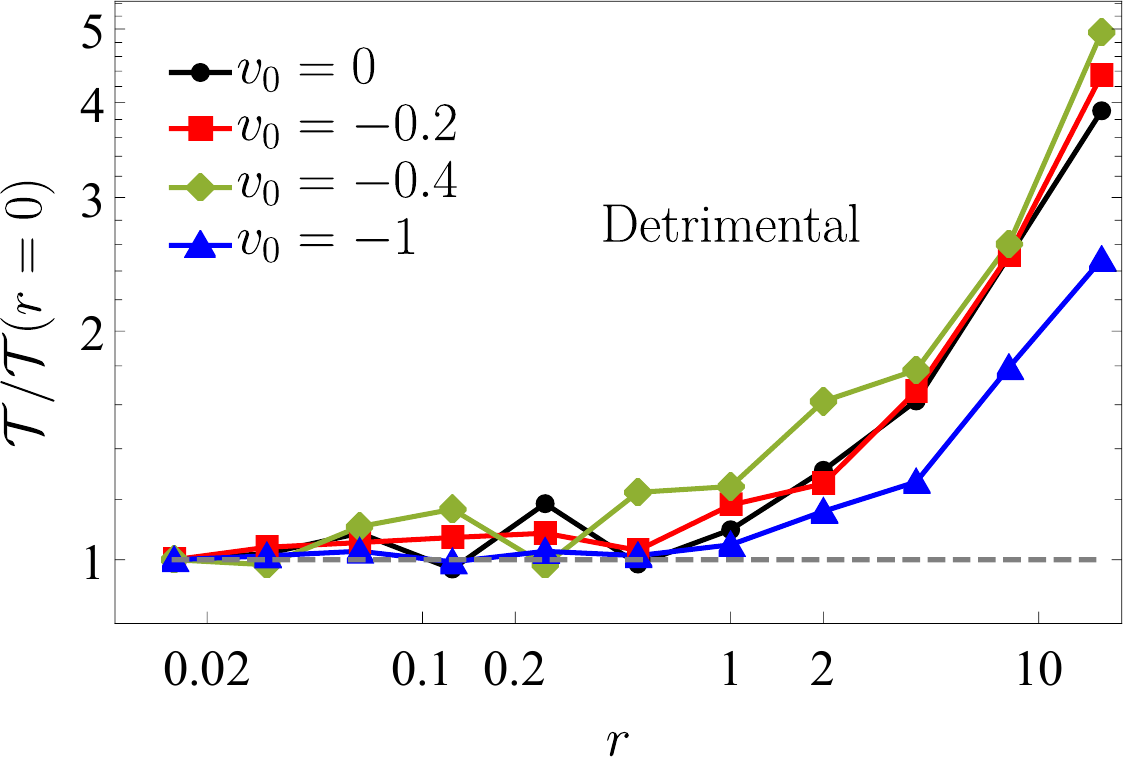}
    \caption{The same as in Fig.~\ref{fig:r-xr0vr1} \bdt{(dependence of the $\mathcal{T}/\mathcal{T}(r=0)$ ratio on the resetting rate)} for position resetting. }
    \label{fig:r-xr1vr0}
\end{figure}

Top \bdt{row (panels (a) and (b))} of Fig.~\ref{fig:CVxr1vr0} shows the dependence of the coefficient of variation, see Eq.~(\ref{eq:cv}), on initial conditions for resetting in position.
The dependence of the $\mathrm{CV}(x_0,v_0)$ is very similar to the one recorded in \bdt{Figs.~\ref{fig:CVxr0vr1}(a) and \ref{fig:CVxr0vr1}(b)}.
For the position resetting coefficient of variation can be significantly larger than 1, i.e., $\mathrm{CV} \gg 1$.
Nevertheless, typically large values of $\mathrm{CV}$ are not associated with acceleration of the escape kinetics due to resetting, see \bdt{bottom row (panels (c) and (d)) of Fig.~\ref{fig:CVxr1vr0}}.
The information provided by $\mathrm{CV}$ is very limited because \bdt{restarting protocol is of non-renewal type}.
The closer examination of individual $\mathcal{T}(r)$ curves indicates that MFPT (also $|x_0| \lessapprox 1$ and $|v_0|\approx 0.5$) is a growing function of resetting rate $r$ and fluctuations of $\Lambda$ are due to numerical accuracy.
As it is visible from Fig.~\ref{fig:r-xr1vr0} the position resetting slows down the escape from the domain of motion.

%
\subsection{Restarting position and velocity \label{sec:resBoth}}

\begin{figure}[H]
    \centering
    \includegraphics[width=1\textwidth]{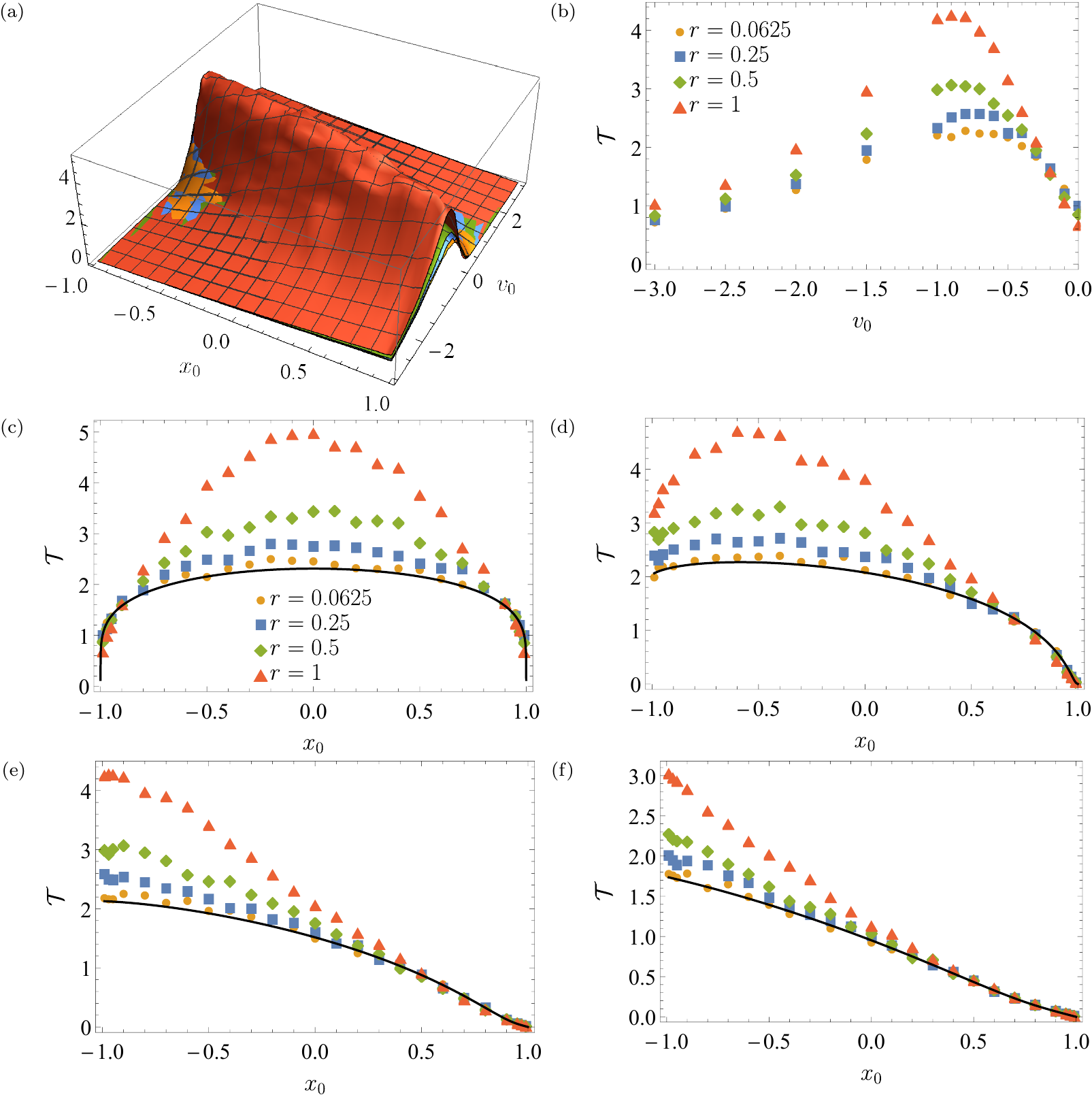}
    \caption{The same as in Fig.~\ref{fig:resVelocity} \bdt{(dependence of the MFPT on initial conditions and resetting rate)} for position resetting.}
    \label{fig:resBoth}
\end{figure}

Fig.~\ref{fig:resBoth} shows dependence of the mean first passage time $\mathcal{T}(x_0,v_0)$ on initial conditions along with sample cross-sections for the last resetting protocol, i.e., for the simultaneous restarting of position and velocity.
\bdt{Among all considered restarting protocols it is the only one which is of the renewal type.}
This restarting protocol bears some of the properties of both resetting protocols.
On the one hand, it prevents unlimited growth of the position distribution and it is capable of trapping the particle inside the domain of motion.
On the other hand, for sufficiently large $|v_0|$ it can be used to maintain particle velocity.
Nevertheless, the role played by the initial velocity is diminished by resetting position back to $x_0$.

\bdt{The inertial motion under random restart, including motion on the half-line, has been studied in \cite{singh2020random}.
The escape from the (positive) half-line, under full resetting, can be characterized by the finite MFPT.
The relevant formula for the MFPT has been derived in \cite{singh2020random}.
Within our studies we explore escape from the finite interval.
Nevertheless, for an appropriate combination of initial condition and resetting rate the escape from a finite interval takes place through one of absorbing boundaries only, see bottom row of Fig.~\ref{fig:semiline}.
In such situations, the escape from the finite interval is equivalent to the escape from the half-line.
Consequently, the MFPT follows predictions of Eq.~(58) from Ref.~\cite{singh2020random}, see Fig.~\ref{fig:semiline} and the detailed discussion in~\ref{sec:app}.
}

The \bdt{full} resetting (simultaneous restarting of the position and velocity) is the most severe resetting protocol, see Figs.~\ref{fig:resVelocity}, \ref{fig:resPosition} and~\ref{fig:resBoth}.
In this scenario, if a particle does not manage to leave the interval before the reset, it starts from a new, i.e., it is moved back to the initial condition \bdt{(renewal affects all the degrees of freedom).}
This in turn significantly decreases chances of removing suboptimal trajectories.

\begin{figure}[H]
    \centering
    \includegraphics[width=1\textwidth]{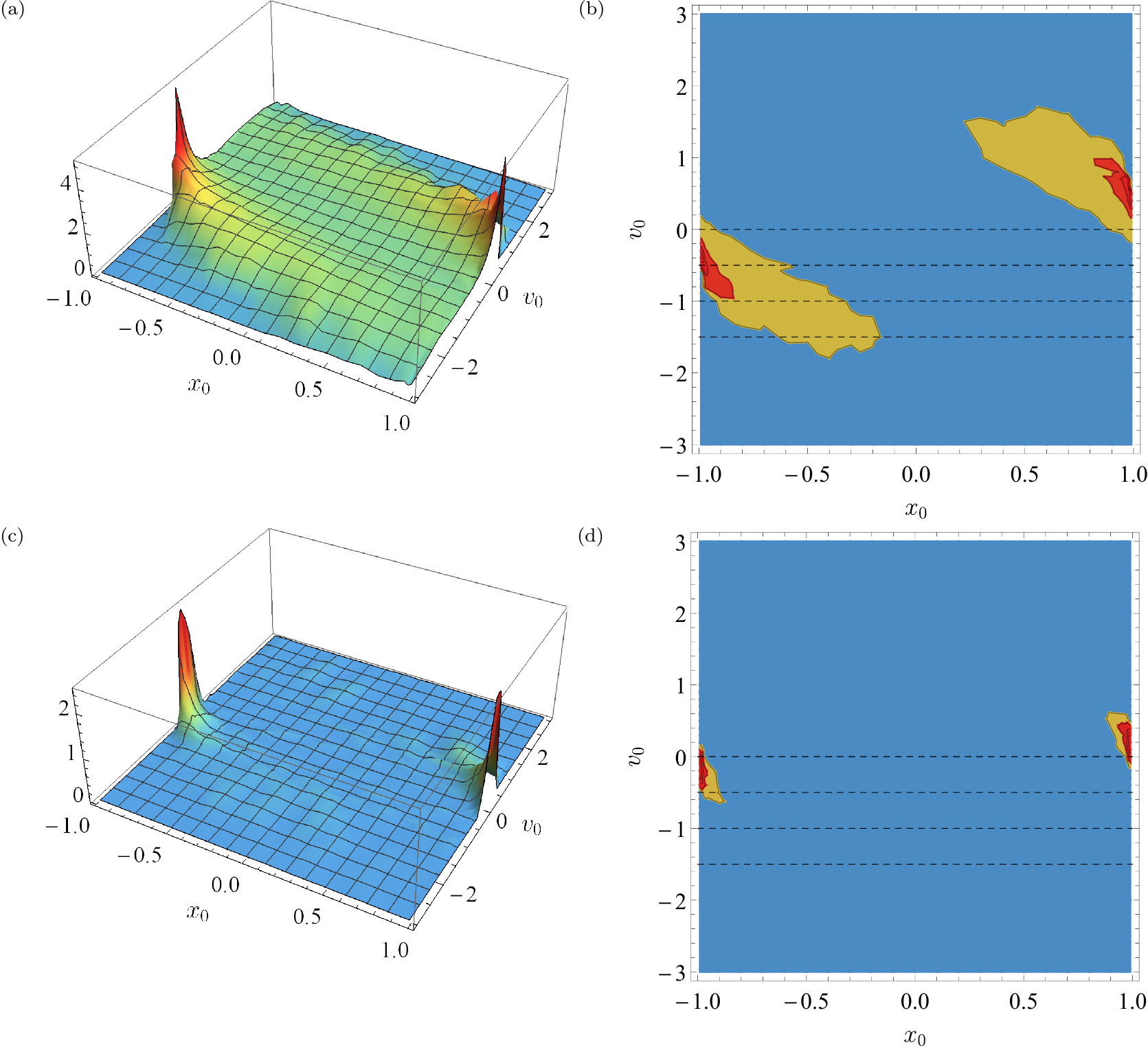}
    \caption{The same as in Fig.~\ref{fig:CVxr0vr1} \bdt{(dependence of $\mathrm{CV}$ (top row) and $\Lambda$ (bottom row) on initial conditions)} for simultaneous position and velocity restarting.}
    \label{fig:CVxr1vr1}
\end{figure}

\begin{figure}[H]
    \centering
    \includegraphics[width=0.7\textwidth]{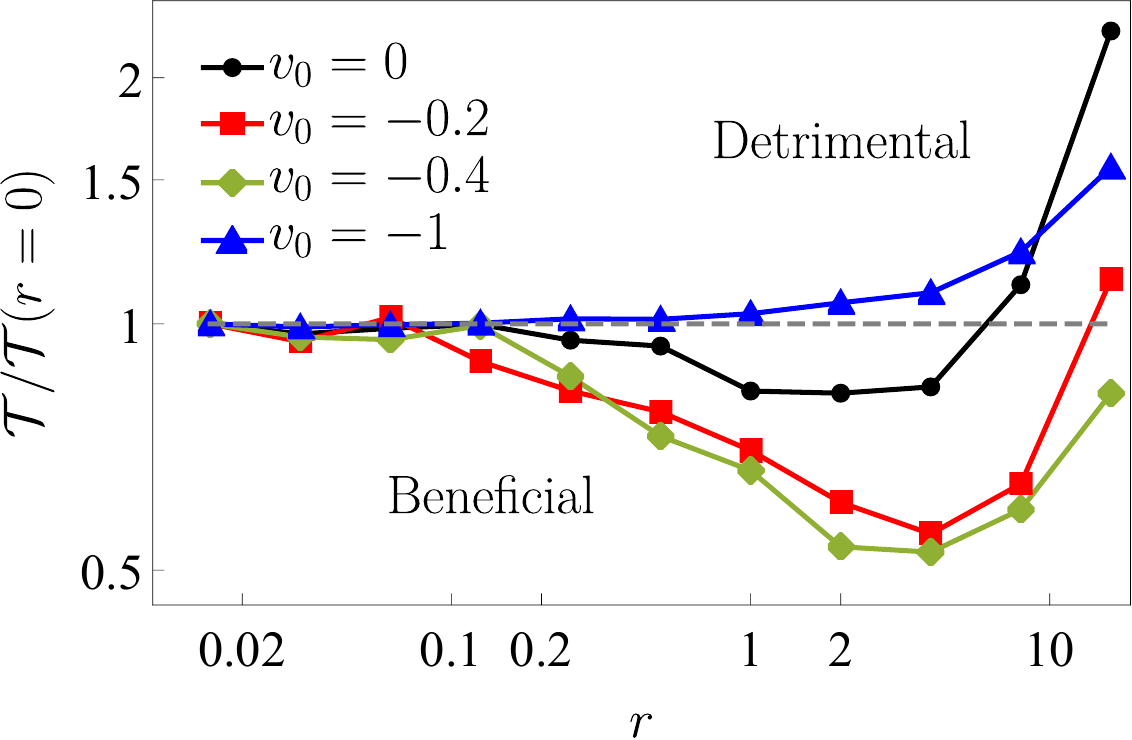}
    \caption{The same as in Fig.~\ref{fig:r-xr0vr1} \bdt{(dependence of the $\mathcal{T}/\mathcal{T}(r=0)$ ratio on the resetting rate)} for simultaneous position and velocity restarting.  }
    \label{fig:r-xr1vr1}
\end{figure}

Top \bdt{row (panels (a) and (b))} of Fig.~\ref{fig:CVxr1vr1} shows the dependence of the coefficient of variation, see Eq.~(\ref{eq:cv}), on initial conditions for simultaneous resetting in position and velocity.
This time the $\mathrm{CV}(x_0,v_0)$ surface manifests the similar behavior like for the velocity resetting (Figs.~\ref{fig:CVxr0vr1}(a) and~\ref{fig:CVxr0vr1}(b)) and position resetting  (Figs.~\ref{fig:CVxr1vr0}(a) and~\ref{fig:CVxr1vr0}(b)).
Nevertheless, comparison of coefficient of variation with the resetting efficiency indicates that the condition $\mathrm{CV}>1$ can be used to localize the domain where resetting expedite escape ($\Lambda>0$) from the finite interval.
This time, in contrast to position resetting, the stochastic resetting indeed can be used to optimize the escape kinetics, see Fig.~\ref{fig:r-xr1vr1}, \bdt{because the escape kinetics can be facilitated for finite, nonzero $r$}.
For example for $x_0=-0.95$ with $0>v_0>-1$ the ratio $\mathcal{R}=\mathcal {T}/\mathcal{T}(r=0)$ is the non-monotonic function of the resetting rate $r$, i.e., for given initial conditions there exist \bdt{a finite} reset rate assuring the fastest escape.
In this case, the information provided by $\mathrm{CV}$ examination is valuable, because \bdt{the renewal
of all the degrees of freedom is preserved}.
Moreover, the motion in a loose sense resembles the overdamped setup.
On the one hand, the position is restarted. On the other hand, velocity resetting decreases the width of the velocity distribution.

\section{Summary and Conclusions\label{sec:summary}}

We have studied the escape of the randomly accelerated, undamped free particle (integrated Brownian motion) from the finite interval accompanied by \bdt{partial or full} stochastic resetting.
We have considered three resetting scenarios: (i) velocity restarting, (ii) position resetting and (iii) simultaneous restart of velocity and position.
\bdt{The resetting protocols (i) and (ii) correspond to the so-called non-renewal (partial) resetting \cite{bodrova2019nonrenewal}, while the last scenario of full resetting is of the renewal type \cite{bodrova2019scaled}.}
The velocity restarting with $v_0\neq 0$ is not capable of trapping the particle in the domain of motion for any value of the resetting rate.
In the limits of $r\to\infty$ the particle performs the ballistic motion.
The particle can be trapped if the velocity is reset to $v_0=0$ or the resetting protocol returns the position to the initial location $x_0$ \bdt{from the interval}.
\bdt{Key results of our studies are summarized in Tab.~\ref{tab:summary}.}

In general, the studied resetting protocols can be ranked by their severity measured by the ability to increase the mean first passage time.
Assuming that the resetting rate is not too large, starting from the mildest one has: position resetting, velocity resetting and simultaneous resetting of position and velocity.
In the limit of infinite resetting rate, if the position is reset, the particle does not leave the domain of motion.
Consequently, for large enough resetting rates, position resetting makes the resetting protocol harsher.
Alternatively, the particle can be trapped under the velocity resetting if the initial velocity is zero.

The coefficient of variation plays the main role determining when stochastic resetting can expedite escape kinetics.
\bdt{For the considered random acceleration process, it has to be analyzed with some care, as not all studied resetting protocols are of renewal type.}
The coefficient of variation surface displays a universal shape which is weakly sensitive to the resetting protocol.
Maximal values of $\mathrm{CV}$ are recorded for initial positions near absorbing boundaries with small initial velocities pointing to the closest absorbing boundary, because, in such a case, there is a striking difference between shortest escape trajectory and non-optimal one.
Since the system dynamics is symmetric with respect to simultaneous exchange of $x_0 \to -x_0$ and $v_0 \to -v_0$, the $\mathrm{CV}$ surfaces display the same symmetry as the mean first passage time, i.e., $\mathrm{CV}(x_0,v_0)=\mathrm{CV}(-x_0,-v_0)$.
\bdt{Under non-renewal resetting,} the coefficient of variation can be used only as a proxy for \bdt{establishing the} domain where resetting could expedite the escape from the interval, because large values of $\mathrm{CV}$ are not necessarily associated with the accelerating of the escape process as it is revealed by examination of the resetting efficiency $\Lambda$.
\bdt{The coefficient of variation provides a robust criterion for the efficiency of stochastic resetting  for resetting protocols securing the renewal of all the degrees of freedom.}
Therefore, it is meaningful for the full (renewal) resetting, i.e., the simultaneous restart of the velocity and position, while for position resetting it does not need to correctly identify the domain where resetting can expedite escape kinetics.

\begin{table}[]
    \centering
    \begin{tabular}{c|c|c|c}
         & velocity resetting & position resetting & velocity and position resetting \\
         & Figs.~\ref{fig:resVelocity} -- \ref{fig:r-xr0vr1} & Figs.~\ref{fig:resPosition} -- \ref{fig:r-xr1vr0} &  Figs.~\ref{fig:resBoth} -- \ref{fig:r-xr1vr1} \\
         \hline \hline
         resetting type & non-renewal & non-renewal & renewal \\
         possibility of trapping & $r\to\infty$ with $v_0=0$ & $r\to\infty$ & $r\to\infty$ \\
         optimal $r$ & $r=0$ or $r\to\infty$ & $r=0$ or $r < \infty$ & $r=0$ or $r < \infty$\\
    \end{tabular}
    \caption{Comparison of studied resetting protocols}
    \label{tab:summary}
\end{table}

The escape kinetics can be optimized by the stochastic resetting of the velocity or by the simultaneous restarting of the position and the velocity.
Both protocols can decrease the mean first passage time, but their modus operandi are very different.
If velocity restarting expedites the escape kinetics it happens for frequent restarts.
Frequent restarting maintains the constant direction of motion by decreasing the width of velocity distribution and keeping it centered at the initial velocity.
\bdt{Under velocity resetting characterized by the large restart rate}, the particle moves ballistically to the boundary indicated by the initial velocity.
The very different situation is observed for the simultaneous position and velocity restarting.
In this scenario, the beneficial role of velocity restarting is counterbalanced by the position resetting.
For appropriately selected initial conditions resetting produces non-monotonous dependence of the mean first passage time on the resetting rate $r$.
Consequently, there exists the optimal restarting rate which expedites the escape from the domain of motion.
The velocity resetting and simultaneous position and velocity restarting should be contrasted with the position resetting, which results in the slowing down of the escape kinetics.

%
%

\appendix
\section{Semi-infinite (half-line) limit\label{sec:app}}

\kct{For the random acceleration process on the half-line the formula for the MFPT under full (renewal) resetting (simultaneous restarting of velocity and position) was recently derived, see \cite[Eq.~(58)]{singh2020random}.
For the clarity of presentation, we rewrite the MFPT formula from Ref.~\cite{singh2020random}
\begin{equation}
    \mathcal{T}_r(x_0,v_0)=-\frac{1}{r}+\frac{1}{r^2\alpha(r)},
    \label{eq:mfpt}
\end{equation}
with
\begin{equation}
    \alpha(s)=\int_0^\infty \frac{dy}{y^{5/3}} e^{-y x_0} \mathrm{Ai}\left(  v_0y^{1/3} + \frac{s}{y^{2/3}} \right) \left[ 1+ \frac{1}{4\sqrt{\pi}} \Gamma\left( -\frac{1}{2}, \frac{2s^{3/2}}{3y}  \right)  \right],
\end{equation}
where $\mathrm{Ai}(\cdot)$ is the Airy function and $\Gamma(\cdot,\cdot)$ stands for the incomplete gamma function.
The escape from the positive half-line can be approached as a special limit of the studied setup.
Within current studies we have used the dimensionless variables, see Eqs.~(\ref{eq:transformation}) and~(\ref{eq:t}), while in \cite{singh2020random} different units are used.
Therefore, in order to compare results of \cite{singh2020random} with outcome of our simulations, the change of units needs to be properly accounted for.
For clarity, we return to the notion with tildes for dimensionless units and without tildes for dimensional units.}

\begin{figure}[H]
    \centering
    \includegraphics[width=1\textwidth]{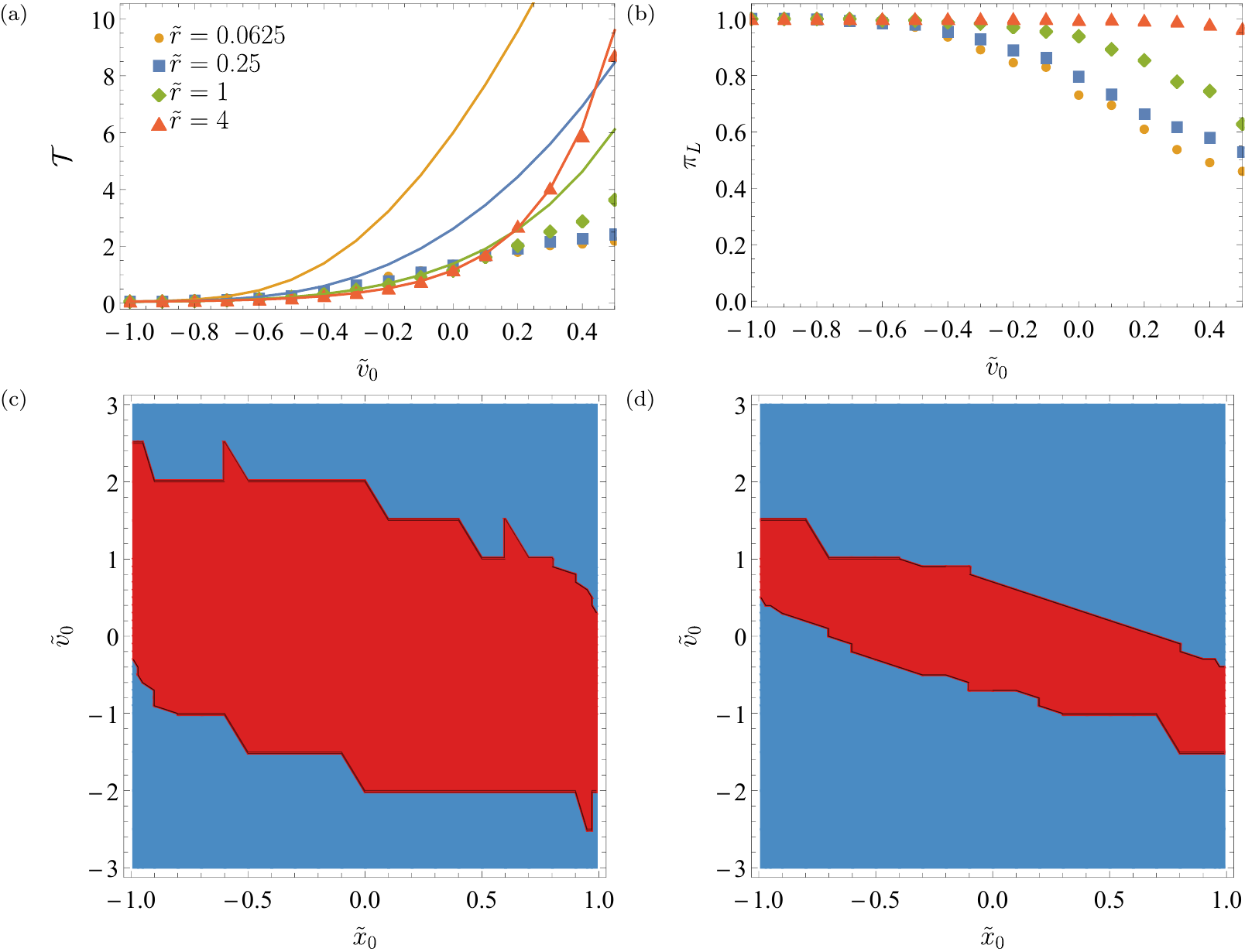}
    \caption{\kct{Top row, in panel (a), presents the numerically estimated MFPT (points) with the appropriately transformed MFPT (solid lines, see Eq.~(\ref{eq:mfpt})), while the panel (b) shows the probability $\pi_L$ of leaving the domain $(-1,1)$ through the left boundary. The initial position is set to $\tilde{x}_0=-0.95$.
    Bottom row depicts regions (colored in blue) where the escape from the finite interval attains the half-line asymptotics for $\tilde{r}=0.625$ (panel (c)) and $\tilde{r}=1$ (panel (d)). 
    The half-line asymptotics is reached if the particle escapes the domain of motion  exclusively through one of absorbing boundaries. }
    }
    \label{fig:semiline}
\end{figure}

\kct{Under full restarting, the half-line limit (in dimensionless units) is reached if all the escapes take place through one boundary, i.e., in the situation when the particle does not ``feel'' the second absorbing boundary.
In order to compare results of simulation with theoretical value of the MFPT given by Eq.~(58)~of Ref.~\cite{singh2020random}, one needs to use Eq.~(\ref{eq:mfpt}) with appropriately selected $x_0$, $v_0=\tilde{v}_0 \times l/T$ and $r=\tilde{r}/T$.
For the initial position one can use $x_0=l-\tilde{x}_0 \times l$ (escape via the right absorbing boundary) or $x_0=\tilde{x}_0 \times l + l$ (escape via the left absorbing boundary), because in Eq.~(\ref{eq:mfpt}) $x_0$ is the distance from the absorbing boundary located at $x=0$.
Since simulations have been performed in the dimensionless variables in transformation of initial conditions one uses $l=1$ and $m=1$.
Furthermore, to reconcile the Langevin equations, the noise strength $\sigma$ needs to be adjusted to $\sigma=\sqrt{2}$.
Finally, the obtained value needs to be divided by $T$.
The described procedure can be successfully used to demonstrate that for an appropriately selected combination of initial conditions and resetting rates, see Fig.~\ref{fig:semiline}, the MFPT attains the half-line limit.
In Fig.~\ref{fig:semiline}(a) points represent results of computer simulations, see Fig.~\ref{fig:resBoth}, while solid lines are the appropriately transformed MFPT given by Eq.~(\ref{eq:mfpt}), see Ref.~\cite{singh2020random} and the discussion above.
The initial position is set to $\tilde{x}_0=-0.95$, thus negative initial velocities facilitate motion towards the left absorbing boundary.
Various curves correspond to various resetting rates $\tilde{r}$.
With the increasing resetting rate $\tilde{r}$ the domain of velocities for which the half-line asymptotic is approached increases.
Numerically calculated MFPTs agrees with the prediction of Ref.~\cite{singh2020random} for the motion on the half-line in the regime where practically all escapes are performed via the left boundary, see Fig.~\ref{fig:semiline}(b) which shows the probability ($\pi_L$) of leaving the interval $(-1,1)$ to the left.
In general, the agreement between the half-line dynamics and the finite interval dynamics is recorded when practically all particles leave the domain of motion to the right or to the left.
The bottom row of Fig.~\ref{fig:semiline} shows the domain where 95\% of escape events are over one of boundaries, i.e., $\pi_L < 0.05$ or $\pi_L>0.95$,  for $\tilde{r}=0.0625=2^{-4}$ (Fig.~\ref{fig:semiline}(c)) and $\tilde{r}=4=2^2$ (Fig.~\ref{fig:semiline}(d)).
Comparison of Fig.~\ref{fig:semiline}(c)) and Fig.~\ref{fig:semiline}(d)) confirms that with the increasing reset rate the domain of validity of the half-line approximations enlarges.}

\kct{Alternatively, one can study escape from the $[0,2l]$ interval in dimensional units, with the fixed initial condition, for increasing the interval half-width $l$, see Refs.~\cite{dybiec2006,dybiec2016jpa}.
For full resetting, such an approach allows for comparison with the exact formula or for finding a large enough $l$ that MFPT stagnates.}
\bdt{
The $[0,2l]$ setup can be also easily used for exploration of other resetting protocols also in situations when MFPT diverges (due to exploration of distant points), for which it is not possible to find finite interval -- half-line equivalence.
In such a case there exist a large, but finite, $l$ for which half-line form of the survival probability is attained.}


\section*{Acknowledgements}

This research was supported in part by PLGrid Infrastructure and by the National Science Center (Poland) grant 2018/31/N/ST2/00598.

%
%
\section*{Data availability}
The data that support the findings of this study are available from the corresponding author (KC) upon reasonable request.

\def\url#1{}
\providecommand{\newblock}{}

\end{document}